\documentclass[11pt]{article}

\usepackage{latexsym}
\usepackage{amssymb}
\usepackage{amsmath}

\numberwithin{equation}{section}

\def\Cl{{\cal C}\ell}
\def\CC{\Bbb{C}}
\def\bz{{\bf z}}
\def\bp{{\bf p}}
\def\bF{{\bf F}}
\def\End{{\rm End}\ }

\begin{document}

\centerline{\Large \bf From Pure Spinor Geometry to Quantum Physics:}
\centerline{\large \bf A Mathematical Way}
\vspace{1.3cm}

\centerline{Paolo Budinich}
\centerline{International School for
Advanced Studies, Trieste, Italy}
\centerline{e-mail: fit@ictp.it}
\vspace{1cm}

\section*{Summary}

In the search of a mathematical basis for quantum mechanics, in order to render it self-consistent and rationally understandable, we find that the best approach is to adopt \'E. Cartan's way for discovering spinors; that is to start from 3-dimensional null vectors and then show how they may be represented by two dimensional spinors. We have now only to go along this path, however in the opposite direction; with these spinors (which are pure) construct bilinearly null vectors: and we find that they naturally generate null vectors of Minkowski momentum space, where the Cartan equations defining pure spinors are identical to all equations of motion for massless systems: both the quantum (Weyl's) and the classical ones (Maxwell's), are determined by them.  These equations are conformal covariance, from which we get, in the conformally compactified phase space, a self dual torus, from which we may rigorously derive the atomic time units like $\Delta t = h/Mc^2$, which P.A.M. Dirac introduced in 1938, as a sign of a ``deep connection in Nature between cosmology and atomic physics''. We have then the possibility of a new, purely mathematical, determination of h: the Planck constant, and thus the possible mathematical starting point for the representation of quantum mechanics.

A fundamental property of pure spinors is that of generating bilinearly null vectors in momentum spaces which, if lorentzian, defines Poincar\'e invariant spheres where quantum dynamical problems should be formulated and solved. This was anticipated in 1935 by V. Fock, who formulated on the sphere $S^3$, the one point compactification of ordinary 3-dimensional momentum space, the non-relativistic integral equation for the H-atom stationary states. He solved it and pointed out the great geometrical visibility of quantization, since the discrete values of the H-atom energy levels $E_n$ are simply due to the, obviously discrete, eigenvibrations of $S^3$ (in contrast to the cumbersome historical approach). We show that the H-atom sphere $S^3$ must be one of those Poincar\'e invariant spheres generated by pure spinors, and in fact $En = - \frac{\alpha^2}{2}\frac{mc^2}{n^2}, n= 1,2,3,\dots $, of obvious relativistic form.

In general pure spinors are also at the origin of the algebraic explanation of the internal symmetry of the ``standard model'' in elementary particle theory. They are also at the mathematical origin of strings which substitute, in quantum mechanics, the concept of point event.

Pure spinor geometry not only appears as the only possible mathematical way to render self-consistent and rationally understandable quantum mechanics but also as a way to define the compact manifolds at the possible origin, and hopefully for the computation, of the several (more than 20) quantum constants, at present inserted by hand, in elementary particle theory.

\section{Foreword}
During the last century, both great revolutions: relativity and quantum mechanics, brought to physics new concepts which are difficult to grasp for our ordinary intuition. Examples are: proper time; for relativity and the collapse of the wave function in space time; for wave mechanics. However while in relativity most of the new concepts are rationally understandable and then acceptable, in quantum mechanics some of them are not so; and 
consequently give rise to surviving paradoxes (like the humoristic ``Einstein's bed")\footnote{A. Einstein was known for telling friends, in Princeton, how much he was disturbed in the evening, when going to bed, by the thought that the wave functions of the constituents of the atoms of his bed were filling his whole bedroom, up to the moment he entered, when they collapsed in the usual corner where he found his bed.}.

Presumably the main motivation of this diversity is that relativity, thanks to Einstein and Poincar\'e, was discovered following a mainly mathematical way, while quantum mechanics followed rather the canonical Galilei's prescription: after the observation and measurement of specific phenomena, formulate the relative physical laws in mathematical language, (in this case the phenomena were: optical atomic spectra and the wave properties of electron beams). The necessity of a mathematical way for quantum mechanics was clearly formulated by Dirac [1] and gave origin to the ``Dirac methodological revolution" [2].

We wish to show here how a mathematical way might be found for the formulation of quantum mechanics if we start from the following motivations.

\section{Epistemological motivations}
\begin{itemize}
\item[a)] The perfect geometrical instrument for the description of classical mechanics of macroscopic bodies; like celestial mechanics, is euclidean geometry of which also the concept of point-event has its role in so far it may represent the centre of mass of the body while describing its trajectory or orbit; as in Kepler celestial motions.

\item[b)] The elementary constituents of macroscopic bodies are fermions, whose relativistic quantum mechanics is well represented with spinors whose geometry is well correlated with  
the physical properties of fermions in so far it geometrically explains, among others, their intrinsic angular momentum or spin.

\item[c)] Spinor geometry was discovered by \'E. Cartan [3] when studying the properties of null (zero length) euclidean vectors. He specially stressed the fundamental role of the spinors he named simple, later renamed pure by Chevalley [4].

\item[d)] Since from the knowledge of the quantum mechanics of the constituent fermions we must pretend to be able to arrive to the knowledge of the classical mechanics of the macroscopic bodies, to reach this goal we should be able to generate with spinors, necessary to represent the quantum mechanics of fermions, the euclidean (and pseudo euclidean) geometry necessary for the description of classical (and also relativistic) mechanics of macroscopic bodies. This will certainly be a mathematical genesis and furthermore the spinorial mathematical object from which we start already contains some quantum features. Of these the main emerged already when P.A.M. Dirac adopted it for a relativistic generalization of the Schr\"odinger equation for the electron (or as a linearization of the Klein-Gordon equation). In fact the four components of his spinor already contained the prediction of a new purely quantum phenomenon: since, of the four components, if two represent the spin of the electron, the other two represent the spin of the positron or anti-electron, discovered by Anderson only 3 years after the formulation of the Dirac's spinor equation, which then predicted quantum phenomena like the creation of the anti-electron, representing thus a perfect example of the Dirac's methodological revolution considered in~[2].
\end{itemize}

Then one may well conjecture that the clue for the mathematical explanation of quantum mechanics may only be found in analyzing the geometrical role of pure spinors when conceived as the elementary constituents of euclidean geometry in a parallel way as fermions are notoriously the elementary constituents of macroscopic matter.

An appropriate instrument for dealing with spinors is represented by Clifford algebras [5]. We will present here a short review of the needed formalism.

\section{Clifford algebras and pure spinors}
Clifford algebras represent an algebraic generalization of the
concept of vector spaces with well defined quadratic forms. We will
start from the case of complex euclidean vector spaces.

Let $W = C^{2n}$ represent a $2n$ dimensional complex euclidean vector
space. Then its associated Clifford algebra $\Cl(2n)$ may be defined
by:
\begin{equation}
\Cl(2n) =  \sum^{2n}_{j=0}\ T_j
\end{equation}
where
\begin{eqnarray}
T_0 &=& \CC   \nonumber\\
T_1 &=& z_a \gamma^a \leftrightarrow \bz \in W\quad a=1,2,3,\dots 2n  \nonumber\\ 
\dots &&\nonumber\\
T_j &=& z_{a_1} z_{a_2} \dots z_{a_j}[\gamma^{a_1},  \gamma^{a_2},\dots \gamma^{a_j}] \nonumber \\
\dots&& 
\end{eqnarray}
where $z_a$ are the $2n$ orthonormal components of a vector $\bz \in W$ and  $\gamma_a$, called the generators of $\Cl(2n)$, obey the anti commutation
relations.
\begin{equation}
\frac{1}{2} [\gamma_a,\gamma_b]_+=\frac{1}{2}(\gamma_a\gamma_b+\gamma_b\gamma_a) = \delta_{ab},\quad a,b=1,2,\dots 2n
\end{equation}
they represent the unit vectors of an orthonormal Cartesian
reference frame in $W$.

We see from (3.1) and the following that in $\Cl(2n)$ we have besides
the field of numbers $\CC$, represented by $T_0$, the vector space $W$,
represented by $T_1$ and all the emisymmetric tensors $T_j$ in $W$.

The elements $T_j$ of $\Cl(2n)$ given by eq.(3.2) may be considered as
operators in a well-defined $2^n$ dimensional space $S$ whose vectors $\psi$ are
spinors associated with $\Cl(2n)$, and then we may write [5]:
\begin{equation}
\Cl(2n) = \End S, \ \  {\rm and} \ \  \psi\in S,
\end{equation}
where $\End S$ stands for endomorphism space of $S$.

For  $\psi\in S$ we have the Cartan's equation
\begin{equation}
T_1 \psi = z^a \gamma_a\psi  = 0 ,
\end{equation}
often the spinor $\psi$ satisfying
eq.(3.5) is called Dirac spinor and indicated with  $\psi_D$. If we
iterate the operator $z^a \gamma_a$ in eq.(3.5) we obtain, because of eq.(3.3); $\bz^2\psi  =0$ and if we adopt for it another vector $\bz '\in W$ different
from $\bz$ we may obtain $\bz\cdot\bz '\psi =0$. This means that a spinor $\psi\in S$ defines through eq.(3.5) a subspace of $W$ whose vectors are all null and
mutually orthogonal, we will call it, in the following, the totally
null plane corresponding to  $\psi_D$ and indicate it with $T_d (\psi_D)$ if $d$ is
its dimension.

We may also define the so-called volume element
\begin{equation}
\gamma_{2n+1} : =  \gamma_1 \gamma_2 \dots  \gamma_{2n}
\end{equation}
which anticommutes with all  $\gamma_a$ and with them generates
$\Cl(2n+1)$ whose even subalgebra $\Cl_0(2n+1)$ is isomorphic to the simple
algebra $\Cl(2n)$, the associated spinors are often named Pauli for
$\Cl_0(2n+1)$ and indicated with $\psi_P$. We also have the Weyl spinors $\varphi^\pm_W$ defined by:
\begin{equation}
\varphi^{\pm}_W=\frac{1}{2}\ (1\pm\gamma_{2n+1})\psi_D
\end{equation}
which are associated with $\Cl_0(2n)$, even subalgebra of $Cl(2n)$. They
satisfy the Cartan-Weyl equations:
\begin{equation}
z^a \gamma_a (1 \pm \gamma_{2n+1})\psi_D = 0\ .
\end{equation}

The totally null planes defined by $\varphi_W$ and eq.(3.8) will be indicated with $T_d (\varphi^±_W)$, if of dimension $d$. It is easily seen that $n$ is the maximal value of $d$, in which case \'E. Cartan named the corresponding spinor simple because of its elegant and simple geometrical properties: $T_n (\varphi^±_W)$ identifies, up to a $+$ or $-$ sign, with the spinor itself [5]. At present such spinors are known as pure, as renamed by Chevalley [4].

Let us now define the main antiautomorphism $B$ of $\Cl(2n) = \End S$ by
$$B: S \to S^\ast$$
where $S^\ast$ is dual of $S$ such that, if $\psi\in S$ and  $\varphi\in S$
\begin{equation}
\langle\psi^\ast /\varphi\rangle = \langle B\psi ,\varphi\rangle\in \CC
\end{equation}
it defines a sort of scalar product.

We may represent the generators $\gamma_a$ in spinor space $S$ with $2^n \times 2^n$
matrices and the spinors with one column $2^n$ component matrices. Then we must have
for $\psi\in S$:
$$B\psi =  \psi^t B\quad {\rm  and}\quad B\gamma_a = \gamma^t_a B$$
where  $\psi^t$ and  $\gamma^t_a$ mean $\psi$ and $\gamma_a$ transposed. It may  easily be seen
that for $\psi$  and $\varphi\in S$ we have [6]:
$$\psi\otimes B \varphi  = \Cl(2n)$$
and then eq.(3.1) becomes
\begin{equation}
\psi\otimes B\varphi = \sum^{2n}_{j=0}\ T_j (\psi ,\varphi )\ , 
\end{equation}
that is all the elements $T_j$ of the Clifford algebra are bilinear
functions of the spinors $\varphi$  and $\psi$. This is precisely what we wanted: a
formalism to rigorously express euclidean geometry elements, in
particular vectors, in terms of spinors. We have not only this but
also the possibility to characterize the spinors which Cartan named
simple and Chevalley pure, setting in clear evidence their
simplicity.

\section{Pure spinors are simple}
\'E. Cartan [3] discovered the spinors he classified as simple,
studying the totally null plains $T_a(\psi_W)$ defined by Cartan-Weyl eq.(3.8) and proved them equivalent, up to a sign, to those null
planes of maximal dimension: that is n. This
definition, which was taken over by Chevalley [4], who named them
pure, implies that, since the dimension of the planes increase
linearly with $n$ while those of the spinors increase like $2^n$ , after
a certain minimal $n(=3)$ the spinors to be pure have to be subject to
a number of constraint equations (bilinear in their components)
precisely $1, 10, 64 \dots$ for $n = 4, 5,6\dots$ respectively, a fact which
will certainly hide their simplicity.

Our approach instead allows to keep it partially in evidence. In fact,
starting from eq.(3.10), let us write Cartan's eq.(3.5) in the
form
\begin{equation}
T_1(\varphi ,\psi )\psi = \langle B\psi , \gamma_a\varphi\rangle\gamma^a\psi =0
\end{equation}
we have:\\

\noindent{\bf Proposition 1.} In a complex $2n$-dimensional
euclidean space $W = C^{2n}$ with associated Clifford algebra $\Cl(2n) =
\End S$, generated by $\gamma_a (a = 1, 2, \dots 2n)$ the vector $\bz\in W$ with
orthonormal components:
$$z_a = \langle B\varphi , \gamma_a \psi \rangle  ,\quad a =1,2,\dots 2n$$
with $\varphi$  and  $\psi\in S$ is null: $z_a z^a = 0$, for arbitrary  $\varphi$ (or  $\psi$) if and
only if $\psi$  (or $\varphi$) is pure.

This theorem, whose proof appears in Ref.[6], characterizes the pure
spinors for any dimension $n$, with the general property of nullness, which assigns them the great elegance of projective geometry.

Let us now suppose to have a $2n$ dimensional pseudo euclidean space $V$
with lorentzian signature:
$$V = R^{(2n-1,1)}\quad {\rm or}\quad V = R^{(1,2n-1)}$$
It is
easily shown [5] that the corresponding components $z_a$ of the vectors $\bz$
appearing in eq.(4.1), may be real if of the form:
\begin{equation}
z_a = p_a = \widetilde{\psi} \gamma_a\psi,
\end{equation}
where $\widetilde\psi = \psi^\dagger\gamma_0$ and  $\psi^\dagger$ means
Hermitian conjugate, and  $\gamma_0$ is the time like generator. Then for $\psi$
pure, because of Proposition 1, ($\varphi$ is arbitrary) $p_a$ will be the
components of a null vector. The corresponding momentum space will
then reduce to the light cone; equivalent to the Poincar\'e
invariant mass-sphere:
\begin{equation}
\pm P_\mu P^\mu = M^2_n=P^2_5+P^2_6+\dots P^2_{2n+2},\quad
\mu = 1,2,3,4
\end{equation}
whose radius $M_n$  increases with $n$; that is, with the dimension $2^{n-2}$
of the fermion multiplet we are dealing with [5].\\

\noindent{\bf Remark 1.} Observe that Proposition 1 is fundamental for
our main aim; which is to show the possibility of constructing
bilinearly the components of pseudo euclidean vectors of lorentzian
signature with pure spinors. This possibility was also conceived by
\'E. Cartan (Ref.[3], Vol.2) and occasionally~[7]  it was named as
Cartan's conjecture.\\

\noindent{\bf Remark 2.}  The Poincar\'e invariant spheres, defined
in (4.3) and deriving from Proposition 1, represent an important
suggestion of momentum space manifolds generated by pure spinors with the
possibility of obtaining geometrically, compatibly with pureness constraints, the several discrete
constants we need in quantum dynamics and or field theory.\\

\noindent{\bf Remark 3.} Observe that from this Proposition 1 the
simplicity of pure spinors appears also in its epistemological
meaning since it means  that pure spinors may be thought as
elementary constituents of euclidean geometry however only of that
sector which is named projective geometry; that is the one of null
euclidean or pseudo euclidean vectors which is certainly more elegant and simple than
the familiar metric euclidean geometry. Not only, but being
dilatation invariant and since we wish to arrive
to classical, celestial or even relativistic-cosmological mechanics,
dilatation invariance is a good company, also because it may provide a good motivation to Dirac's conjecture of the ``deep connection in Nature''.

In
order to illustrate transparently this possible mathematical way to
quantum mechanics we will start from the simplest non trivial two
component pure spinors, corresponding to $n=1$. This will also allow to directly verify and control the fundamental importance and role of eq.(3.10).
Since $n=1$, in eq.(4.3) we will have $M^2_1=0$; we may then expect to deal with massless systems.

\section{From pure spinors to quantum mechanics}

\subsection{From $n=1$ to $n=2$. The Minkowski signature, the Weyl and
Maxwell field equations}

Start from $\Cl(2)$ and let $\varphi = \left({\varphi_0\atop\varphi_1}\right)$ and $\psi = \left({\psi_0\atop\psi_1}\right)$ represent two of its
associated Dirac spinors, or Pauli spinors of the isomorphic $\Cl_0(3)$,
generated by the Pauli matrices $\sigma_1, \sigma_2, \sigma_3$. Insert them in equation
(3.10), where now $B = -i\sigma_2 = :\Sigma$, which becomes:
\begin{equation}
\left( {\varphi_0\psi_1 \ \ -\varphi_0\psi_0\atop
\varphi_1\psi_1 \ \ -\varphi_1\psi_0}\right)\equiv \varphi\otimes B\psi =
z_0+z_j\sigma^j\equiv \left( {z_0+z_3 \ \ z_1-iz_2\atop z_1+iz_2 \
z_0-z_3}\right)
\end{equation}
from which we easily get both the $z$-vector
components bilinear in the spinors $\psi$  and $\varphi : z_\mu  = \frac{1}{2}\psi^t
\Sigma\sigma_\mu\varphi$, where  $\sigma_0=1$
(compare the matrices) and the nullness of the vector $\bz : z_\mu  z^\mu =
z^2_0-z^2_1-z^2_2-z^2_3\equiv 0$  (compute the determinants of the matrices)
in agreement with Proposition 1.

In order to restrict to the real $z_0$ and $z_j$, of interest for physics,
we need to introduce the conjugation operator, $C$ defined by: $C\gamma_a  = \bar\gamma_a C, C\varphi  = \bar\varphi C$, where $\bar\gamma_a$ and $\bar\varphi$ mean $\gamma_a$ and $\varphi$ complex conjugate. Then
equation (5.1) may be expressed, and uniquely, in the form:
$$
\left( {\varphi_0\bar\varphi_0 \quad \varphi_0\bar\varphi_1\atop
\varphi_1\bar\varphi_0 \quad \varphi_1\bar\varphi_1}\right) =
p_0+p_j\sigma^j= \left( {p_0+p_3 \ \ p_1-ip_2\atop p_1+ip_2 \
p_0-p_3}\right) \eqno(5.1')
$$
and now:
\begin{equation}
p_\mu  = \varphi^\dagger\sigma_\mu\varphi,\quad  \mu = 0,1,2,3
\end{equation}
where $\varphi^\dagger$  means $\varphi$ Hermitian conjugate. Then we have, again
identically:
\begin{equation}
p_\mu p^\mu  = p^2_0 - p^2_1 - p^2_2 - p^2_3 \equiv  0
\end{equation}
which shows how
$p_\mu$ are the components of a null or optical vector of a momentum space
with Minkowski signature. This is a particular case of application
of Proposition 1. In fact imbed $\Cl_0(3)$ in the non simple $\Cl(3)$
isomorphic to $Cl_0(1,3)$ with generators $\gamma_\mu   =  \left\{ \sigma_1\otimes 1,-
i\sigma_2\otimes\sigma_j\right\}$ and $\gamma_5= -i \gamma_0
\gamma_1 \gamma_2 \gamma_3 = \sigma_3\otimes 1$. Then we may identify the above Pauli spinor
with one of the two Weyl spinors defined by~[5]:
\begin{equation}
\varphi_\pm = 1/2 (1\pm \gamma_5) \psi ,
\end{equation}
where $\psi$  is a Dirac spinor associated with $\Cl(1,3)$. Then equation
(5.2) identifies with one of the two:
\begin{equation}
p^{(\pm )}_\mu  = \widetilde\psi\gamma_\mu (1\pm\gamma_5)\psi  ,\quad
\mu = 0,1,2,3,
\end{equation}
where $\widetilde\psi = \psi^\dagger\gamma_0$.
Now the vectors $\bp^{\pm}$ are null because of Proposition 1, since the Weyl
spinors $\varphi^\pm$  are pure. 

As seen in chapter 3 Weyl spinors obey the Cartan-Weyl equations (3.8) which now will
be:
\begin{equation}
p_\mu\gamma^\mu (1\pm\gamma_5)\psi = 0
\end{equation}
which may be expressed in Minkowski space-time if $p_\mu$  are interpreted
as generators of Poincar\'e translations: $p_\mu\to   i
\partial /\partial x_\mu$. They identify,
after introduction of the Planck's constant $h$, with the known wave
equation of motion of massless neutrinos.
$$
ih\frac{\partial}{\partial x_\mu}\gamma^\mu (1\pm\gamma_5)\psi =0\ .\eqno(5.6')
$$
Observe that in this unique
derivation, obtained by merely imposing the reality of the $p_\mu$
components, Minkowski signature naturally derives, and, since
$\Cl(1,3) = H(2)$, where $H$ stands for quaternions, we may affirm that
Minkowski signature is the image in nature of quaternions.

It is interesting to observe that if we define the electromagnetic
(so named already by \'E. Cartan [3]) tensors $\bF$ with components:
\begin{equation}
F^{(\pm )}_{\mu\nu}=\widetilde\psi [\gamma_\mu ,\gamma_\nu ]
(1\pm\gamma_5)\psi
\end{equation}
which define the maximal totally
null plane equivalent, up to a sign, to the pure spinors defined in
eq.(5.4), we obtain, from Cartan-Weyl equations (5.6) the Maxwell's
equations in empty space:
\begin{equation}
p_\mu F^{\mu\nu}_+ = 0,\quad \varepsilon_{\lambda\rho\mu\nu} p^\rho
F^{\mu\nu}_- = 0.
\end{equation}
The electromagnetic tensor is then equivalent (up to a
sign) to massless neutrinos, but notoriously not their bound states [5]. 

In the
last part of this section we naturally operated the transition from
$n = 1$ to $n = 2$, it is easy [5] to generalize it to show how to go in even
dimensional lorentzian spaces from $n$ to $n +1$ to steadily obtain,
from pure spinors, null vectors with real components.

Up to now we arrived through pure spinor geometry to the Weyl
quantum field equations for massless neutrinos and from these to
Maxwell's ones; that is to both quantum and classical equations
for massless systems; not surprising in the projective momentum
space which we constructed with pure two component spinors corresponding to $n=1$.

However we also encounter our first severe difficulty: how to
justify that we need as a factor the Planck's constant $h$ in Weyl
equations (5.6$'$) and not in Maxwell (5.8) ones? And, since we pretend to
follow a mathematical way how can we compute it? It seems a
desperate difficulty. However there might be a positive answer
correlated with the dilatation invariance that we keep with us.\\

\noindent{\bf Remark.}
Observe that in this section we derived from the simplest pure spinor geometry $(n=1)$ the field equations for the simplest physical systems: the massless neutrinos and photons, in vacuum: that is, without interactions. Pure spinor geometry may also give, and in fact gives (for higher $n$), also field equations with interactions. In particular, for the massless systems: neutrinos and photons, the interactions happen to be [5] exactly those of the Electro-Weak model, which in this way results explained by pure spinor geometry.

In general, it happens that, for higher $n$ pure spinor geometry gives exactly all and only those interactions which are observed in nature. This will be briefly summarized in the following section 5.3.

\subsection{The role of the conformal group}
The equations of motion
that we derived are Lorentz covariant but also,
notoriously, covariant with respect to the larger conformal group $C$:
\vspace{0.5cm}
\begin{equation}
C = L \otimes D \times\!\!\!\!\!\supset P_4 \times\!\!\!\!\!\supset S_4
\end{equation}
when $L$ stands for Lorentz, $D$ for dilatations $P_4$ for Poincar\'e
translations and $S_4$ for special conformal transformations,
respectively. For us they are of capital importance since they
contain dilatations.

It is well known that the conformal group may be linearly
represented by the pseudo orthogonal group $SO(4,2)$ acting in the
pseudo euclidean space $W = R^{4,2}$; obviously, the group without
reflections. However we know that reflections are important for
physics (parity violation) therefore we have to take all of them
including the so-called conformal reflections I; that is those with
respect to hyperplanes orthogonal to the 5$^{\rm th}$ and the 6$^{\rm th}$ axis in
$R^{4,2}$. In order to obtain a representation of the conformal group $C$
with reflections we must [8] introduce the following homogeneous spaces:
\vspace{0.5cm}
\begin{equation}
\left.\begin{array}{lr}
& M_c=\frac{C}{c_1}\\
{\rm and:}&\\
&P_c =\frac{C}{c_2} \end{array}
\right\} = \frac{S^1\times S^3}{Z_2}
\end{equation}
where $c_1 = \{ L, D, S_4\}$  and $c_2 = \{ L, D, P_4\}$  are the stability groups
of the origin $(x_\mu  = 0)$ and of infinity respectively, and $Z_2$ means
that two points diametrically opposite with respect to the centers
of the spheres have to be considered equivalent. $M_c$ has often been
considered as a model for the conformally compactified space-time
and precisely for Robertson-Walker Universe now generally adopted in
cosmology. Correspondingly $P_c$ has been taken as a model for
conformally compactified momentum space.

It is known [8] that $M_c$ and $P_c$ transform in each other for conformal
reflections I:
\vspace{0.5cm}
\begin{equation}
M_c = I\ P_c I^{-1}
\end{equation}
because of which we may affirm that $M_c$ and $P_c$ are
conformally dual. We may consider them as two copies of the
homogeneous space for the representation of the conformal group with reflections.

At this point it is appropriate to remind that, in quantum wave mechanics space-time and momentum-energy, spaces are conceived as correlated by Fourier transforms. Then if we reasonably postulate that space-time must be densely contained in $M_c$ as well as momentum-energy densely contained in $P_c$ and if we further recognize as well established by celestial mechanics as well as by general relativity that space-time is well appropriate for the mathematical representation of classical mechanics, then eq.(5.11) might suggest that momentum-energy is the space for the mathematical description of quantum mechanics. This is incoherent with what in fact appears in the beginning of this chapter where we perform the first steps on the mathematical way for constructing the geometry bilinearly generated by pure spinors  starting with the simplest, and we arrive indeed to quantum equations of motion in Minkowski geometry but of momentum space.

Let us now suppose that not only compactified space-time $M_c$ is
realized in nature (Robertson-Walker Universe) but also $P_c$, both of
the form (5.10). Phase space would then be compact and as such of
great interest for quantum field theory, since in principle, one
could hope to get rid of the old unsolved problems of infrared and
ultraviolet divergences [8]. For our purpose it is enough to observe
that the resulting phase space will contain a torus $T_2$.
\begin{equation}
T_2 = S_T^1 \times S_E^1
\end{equation}
in
which the first $S_T^1$  represents time of $M_c$ and the second $S_E^1$  energy
of $P_c$. Now, according to (5.11), for conformal reflections the Torus
in eq.(5.12) remains invariant since $S_T^1$  transforms to $S_E^1$  and
viceversa. Time and energy are complementary with respect to the uncertainty principle and often Fourier dual in quantum physics and
we may well suppose that $S_T^1$  and $S_E^1$  are discrete lattices apt to
determine both an elementary time unit (spacing of the time lattice)
and the value of the Planck's constant $h$.

In fact suppose now to inscribe in $S_T^1$ and $S_E^1$  two regular polygons
with $2N$ vertices,  from the rules of discrete Fourier
transforms, denoting with $\Delta t= \pi T/N$ (where $T$ is the radius of $S_T^1$)
the spacing of the time lattice $S_T^1$, and with $Mc^2$ the radius of
$S_E^1$, you easily obtain  (see Chapter 2: The toy model of $S^1$ in Ref.[8])
\begin{equation}
\Delta  t Mc^2 = h\ .
 \end{equation}
If $M$ is the mass of the proton, we obtain 
\begin{equation}
\Delta  t =h/Mc^2= 4.4 \times 10^{-24}\ \textrm{sec.}
\end{equation} 

\newpage

\noindent
which is one of
the atomic time units considered by Dirac\footnote{P.A.M. Dirac in 1938 defined several atomic time units like $\Delta t$ in eq.(5.14), as well as further atomic parameters (like the ratio of the electric and gravitational forces between electron and proton) presenting large ratios: of the order of $10^{39}$ with cosmological constants, concluding that they may not be accidental but rather the indication of a ``deep connection in Nature between cosmology and atomic theory''. He then continued, during his whole life, to further elaborate this idea. In our approach we could justify his conjecture as deriving from the role of dilatation covariance in the conformal group; which rules the symmetry of the dynamical behaviour of all massless systems both the classical (Maxwell) and the quantum (Weyl) ones.} [9]  because of its ratio
$3.4\times 10^{39}$ with the age of the Universe (which equals $N$ above). It is here determined
geometrically from $S_T^1$ of the Torus $T_2$ in eq.(5.12).

Clearly at this point one should discuss the possibility of discrete
time, which is considered in Ref.[10] as a reasonable possibility,
in the frame of non commutative geometry\footnote{It appears
compatible with Lorentz covariance and correlated with the old idea
of quantized space-time of S. Snyder [11]  and C.N. Yang [12] . It is
interesting to observe that if it happens to be right, then from eq.(5.13) one might conclude that Planck's constant equals the product
of the minimal time interval in the form $h/Mc^2$ with
the rest mass energy of the proton: the only probably stabile and most abundant
baryon of the visible Universe.}.

\subsection{Up to $n=5$}
Next we may consider several components
spinors: say $2^n$ with $n> 2$ representing multiplets of $2^{n-2}$ fermions, to see what they
foresee for the complex phenomena  recently discovered in high
energy elementary particles phenomenology. It was shown in [5]  how
the internal symmetry of the Standard Model;  $[U(1), SU(2)_L, SU(3)]$
may easily be attributed to the role of the $3$ complex
division algebras (complex numbers, quaternions and octonions) in
Clifford Algebras of the form $\Cl(2n-1,1)$ with $n$ going from $n = 1$ to
$n = 5$ (corresponding to $\Cl(9,1)$) and not beyond because of the Bott
periodicity theorem. In this way several as yet obscure aspects of
elementary particle phenomenology, as the electro weak model (see the final Remark in section 5.1), the $3$
families and $3$ colours because of the $3$ imaginary units of
quaternions and so on, were easily explained.

The spinors with which we deal in this paper, in particular the pure ones, are vectors of the endomorphism space of Clifford Algebras, like $\Cl(2n)$ or $\Cl(2n+1)$, and, as such their $2^n$ components are either complex or real numbers, and then commuting (and only as such they may be conceived, say, as elementary constituents of projective geometry).

It is well known that in quantum field theory, spinors have to be considered as operators (building up in particular creation and annihilation operators), which generally are anticommuting (to correctly represent the Pauli exclusion principle). As such, pure spinors have been recently successfully used in the computations of super symmetric quantum gravity. In this paper we will not deal with this certainly interesting and promising aspect of pure spinor geometry.

But the most promising results, for quantum mechanics, arriving from
pure spinor geometry derive from the fact that Proposition 1 defines
compact manifolds in momentum space, and then opens the possibility
of a new formulation and solutions of problems in quantum dynamics, setting in evidence the geometrical origin of the discreteness of certain quantities and the geometrical origin of quanta.

\section{Quantum Dynamics in Compact momentum space}
Both from our
construction of projective geometry from pure spinors and from
Proposition 1, we may draw the suggestion that quantum dynamical
problems should be mathematically formulated and
solved in compact momentum space. An impressive anticipation of this
opportunity may be found in a paper by V. Fock [13]  who, in 1935,
formulated and solved the emblematic problem of the H-atom
stationary states as an integral equation on $S_3$: one
point compactification of ordinary $3$-dimensional momentum space.
With brilliant results in so far he not only showed how the discrete
H-atom energy levels are simply due to the obviously discrete eigen
vibrations of the sphere $S_3$ but also discovered the obvious $SO(4)$
symmetry of the problem formulated on $S_3$, later extended by W. Pauli also to planetary
systems.

Because of Proposition 1, once we use
lorentzian momentum spaces and the spheres of eq.(4.3), we may, for
any $n$, use spinors, and be sure that they are pure without imposing
the constraint relations, provided we are not going out of the manifolds defined by eq.(4.3).  However it may well
be that pureness might impose some geometric-algebraic constraints for action 
on the spheres of eq.(4.3). We have already some for the Fock's
H-atom equation, which may be written in the form:
\begin{equation}
\psi (\mathbf{u}) =\frac{\alpha}{V(S_3)}\frac{mc}{p_0}\int_{S_3} \frac{\psi(\mathbf{u}')}{(\mathbf{u}-\mathbf{u}')^2} d^3\mathbf{u}'
\end{equation}
where $V(S_3) = 2\pi^2$ is the volume of the unit sphere $S_3, \alpha = \frac{e^2}{\hbar c}$  is the fine structure constant, $p_0$ is a unit of momentum, $m$ the electron's mass, and $\mathbf{u}$ is a vector indicating a point on the unit sphere $S_3$. This equation is the one adopted in 1935 by V. Fock [13] (set here in adimensional form) for the description of the H-atom in the one point compactification $S_3$ of ordinary 3-dimensional momentum space. Fock showed how this equation solves the problem of H-atom stationary states, after a harmonic analysis from the ball $B_3$ to $S_3$ which, for $\psi\to\psi_n$: spherical harmonics on $S_3$, gives for $E = –p_0^2/2m$ the known eigenvalues $E_n$ of the H-atom stationary states. This solution also sets in evidence the $SO(4)$ symmetry of the H-atom system. We see then that in the spheres in momentum space we may solve purely geometrically the dynamical problem of at least a simple but emblematic system as that of the H-atom, at difference with the cumbersome historical way in space time.

Let us now derive Fock's equation (6.1) from pure spinor geometry.

The sphere $S_3$ may be obtained [7] from eq.(4.3) for $n = 3$; since the H-atom may only be obtained from a doublet containing both the proton and the electron and then eq.(4.3) becomes:
\begin{equation}
\pm P_\mu  P^\mu  = M_3^2 = P_5^2+P_6^2+P_7^2+P_8^2
\end{equation}
which defines $S_3$, however at difference with the Fock's one, this is relativistically invariant, as in fact the Balmer energy levels of
the H-atom are:
\begin{equation}
En = -\frac{\alpha^2}{2} \frac{mc^2}{n^2},\quad n = 1, 2, \dots
\end{equation}
Observe that in Fock's eq.(6.1), in front of the integral, there are two adimensional factors; besides $mc/p_0$ whose value was determined by V. Fock there is also the fine structure constant $\alpha$. We could try to compute it starting from a higher dimensional spinor, that is to take $n = 4$ instead of $3$ which corresponds to $16$ component spinors representing two fermions doublets, which is appropriate since it is known [5] that in nature there are two main classes of fermions: the baryon and the leptons and, in general, as far as the electric charge is concerned, in nature they frequently appear in charged neutral pairs. Therefore to represent 
the Hydrogen atom instead of starting from proton $+$ electron, what we did above with $n = 3$, it is better to start from proton $+$ neutron $+$ neutrino $+$ electron (it may be shown that if you take them in this order and you assign a positive electric charge to the proton the electron naturally results negative [7]), corresponding to $n = 4$ in which case (6.2) becomes:
$$
\pm P_\mu  P^\mu  = M_4^2 = P_5^2+P_6^2+P_7^2+P_8^2+P^2_9+P^2_{10}
\eqno(6.2')$$
representing a Poincar\'e invariant $S_5$ sphere, which contains the invariants $S_4$ and $S_3$.

Therefore one may hope to obtain the value of the adimensional fine structure constant $\alpha$ in eq.(6.1), through harmony analysis, like Fock did it for the factor $mc/p_0$ from eq.(6.1) in $S_3$, but this time in $S_4$ and the correlated classical symmetric domains $D_5$ with boundary $Q_5$. Now it happens that 3 authors [14], [15] and [16], have independently computed it in terms of precisely these domains finding
\begin{equation}
\frac{e^2}{\hbar c}=\frac{8\pi[V(D_5)]^{1/4}}{V(S_4)V(Q_5)}= \frac{1}{137,0608}
\end{equation}
which differs less than $1/10^6$ from the experimental value. Therefore eq.(6.1) could be written in the form:
\begin{equation}
\psi(\mathbf{u})=
\frac{8\pi[V(D_5)]^{1/4}}{V(S_3)V(S_4)V(D_5)} \frac{mc}{p_0} \int_{S_3}
\frac{\chi(\mathbf{u}')}{(\mathbf{u}-\mathbf{u}')^2} d^3\mathbf{u}'
\end{equation}
indicating the possibility of a purely geometrical formulation of the H-atom problem including the computation of $\alpha$. It is obvious that should one be able to compute the value of the fine structure $\alpha$ in eq.(6.1) one could, by reversing the steps which were performed by Fock, arrive at the Schr\"odinger equation in space-time where both the electric charge and Planck's constant would appear with their values geometrically defined. At difference with the traditional approach in which they have to be inserted by hand, thus opening the door to the possibility of a full geometrization of quantum mechanics.

The methods with which the mentioned 3 authors computed the value of $\alpha$ in eq.(6.4) are quite different and difficult to follow and justify. Now since Fock followed the rules of harmonic analysis for computing $mc/p_0$ we (with P. Nurowski) followed the same rules for $\alpha$ starting from $S_4$ and we found all the factors in eq.(6.4) but one $([V(D_5)]^{1/4})$.

It might well be that for acting on the manifolds of eq.(6.2$'$) one has to take into account the constraint relations for pure spinors (of which our vectors are bilinear) which are $1$ and $10$ for $8$ and $16$ component pure spinors, respectively. Alternatively one could modify the rules of computation of harmonic analysis. This will be discussed elsewhere. In any case should a way out be found, then it would allow a first big step forword for the theory of elementary particles since at present in the ``Standard Model" more than $20$ quantum constants like charges and masses have to be inserted by hand.

\section{Conclusions and outlook}

In our search for a rigorous mathematical way for the formulation of quantum mechanics the simplest was to reverse the way found by \'E. Cartan for the discovery of spinors. He started from null vectors (in projective geometry) and discovered spinors as their square roots, so to say. Among these he soon discovered the simple spinors, now renamed pure. Reversing this itinerary it is not surprising that pure spinors may bilinearly generate the null vectors of projective geometry (Proposition 1). What is new, is that in the first steps of our construction with two component pure spinors $(n=1)$, we naturally obtained (in chapter 5) Minkowski signature, however in momentum space, where we also obtain Weyl's and Maxwell's equations, which thus manifest their pure spinorial and then geometrical, origin. To get them all in ordinary space-time we may exploit their Lorentz and conformal covariance for massless quantum systems, which allows the mathematical definition of Planck's constant $h$.

Therefore we may conclude that relativity and quantum mechanics have a unique starting point for their mathematical formulation: the geometry of pure spinors which are the elementary constituents of the projective sector of euclidean geometry.

From this projective sector, we may arrive, through integrations of null vectors, to the familiar metric euclidean geometry, necessary for the representation of classical and relativistic mechanics either through minimal surfaces [17] or strings, depending from the signature. More precisely for Minkowski signature we obtain bilinearly from spinors only strings [18] and we may then interpret this as the necessary geometrical constraint for the mathematical way to quantum field theory in which the concept of point-event is not needed but rather substituted by that of strings as supported by experimental evidence at CERN (Geneva), as then in fact generally adopted by the most distinguished theoretical physicist in quantum field theory.

For relativity from Maxwell's equations one extends their Lorentz covariance to mechanics and then considers the Riemannian structure of space-time to arrive to general relativity and cosmology for which space time remains the ideal arena for the mathematical description of this macro world, as wonderfully shown by Einstein and Poincar\'e.

For quantum mechanics instead, once obtained a mathematical definition of the Planck's constant $h$, as mentioned above, one has to operate in momentum space which, thanks to Proposition 1, gives us the compact manifolds at the origin of the quantum phenomena of the atomic micro world, whose dynamics appears trivial but in the mathematical space of velocities, or momentum space, and if the mathematical way is compatible with spinor pureness, as shown above for the V. Fock example.

This mathematical path through momentum space should be further explored through more problems, and further exploiting the geometry of pure spinors to render rationally understandable and then acceptable also quantum mechanics, which is not, up to now; not only because of the difficulties of quantum gravity, but rather for the large number of quantum constants which could be hopefully explained and also computed in the compact manifolds like those of eq.(4.3) determined by pure spinor geometry in momentum spaces, on which as allowed by Proposition 1 one may act with the only mathematical instruments compatible with spinor pureness.

It might also be helpful to consider these developments in the frame of the Dirac's methodological revolution [2].

We know with mathematical certainty that with pure spinors we may bilinearly generate null vectors, which may define compact manifolds (spheres) in lorentzian momentum spaces where quantum problems may be formulated and easily solved, taking advantage of pure spinor geometry and of some chapters of pure mathematics like harmonic analysis. It is to be expected that this method may be further extended to the recent results in high energy experiments in elementary particle physics: there are several well defined data which up to now have no rational explanation while they clearly refer to quantum problems correlated with multi component spinors and therefore they could be formulated and solved in the compact manifolds mentioned above. Obviously in the mathematical formalism computations with pure spinor geometry, one has to follow strictly the rules of pure spinors geometry and possibly some sophisticated development of harmonic analysis. After all in the spirit of ref. [2] in this paper we have simply given some good arguments on why pure spinors might be conceived as the elementary constituents of euclidean geometry in a parallel way as fermions may be conceived as the elementary constituents of macroscopic bodies, which is a metaphysical result; one has simply to exploit and to extend this knowledge, obtained from this particular sector of mathematical metaphysics [2]. The fact that mathematical metaphysics may bring new knowledge is old and well known in cosmology if one thinks of gravitational lenses, pulsars, black holes and so on.
%\vspace{1cm}

%\noindent{\bf Acknowledgments.}

%The author wishes to express his gratitude to L. Bonora, M. Budinich, P. Furlan, P. %Nurowski, G. Thompson and A. Trautman for helpful advice and discussions.

\end{document}